\documentclass[aps,pra,epsfigure,twocolumn,showpacs]{revtex4}
\usepackage{dcolumn}    
\usepackage{bm} 
\usepackage{graphicx}
\usepackage{amsmath}    
\usepackage{latexsym}
\usepackage{amsfonts}   
\usepackage{amssymb}
\usepackage{array}      
\usepackage{epsfig}

\newcommand{\ket}[1]{\left\vert#1\right\rangle}

\newcommand{\miniproj}[3]{\vert#1\rangle_{#2}\!\langle#3\vert}

\newcommand{\minisand}[3]{\langle#1\vert#2\vert#3\rangle}

\begin{document}
\title{Teleporting bipartite entanglement using maximally entangled mixed channels}
\author{S. Campbell and M. Paternostro}
\affiliation{School of Mathematics and Physics, Queen's University, Belfast BT7 1NN, United Kingdom}

\begin{abstract}
The ability to teleport entanglement through maximally entangled mixed states as defined by concurrence and linear entropy is studied. We show how the teleported entanglement depends on the quality of the quantum channel used, as defined through its entanglement and mixedness, as well as the form of the target state to be teleported. We present new results based on the fidelity of the teleported state as well as an experimental set-up that is immediately implementable with currently available technology.  
\end{abstract}
\date{\today}
 \maketitle

\section{Introduction}
\label{Intro}
Among the attracting, puzzling and intriguing features of quantum mechanics, entanglement certainly occupies a premier position. This explains why the last two decades have seen remarkable efforts produced towards the generation, characterization and exploitation of correlations of truly non-classical nature. Theoretically, it is now a wide-spread belief that entanglement is a valuable resource intackling tasks of information processing. One of the contexts where the pragmatic potential offered by entanglement appears to be magnified turns out to be the dynamics of {\it delocalized} networks of ``quantum'' nodes, designed in order to accomplish a task. If such nodes share entangled channels for communication and computation, with the complement of ``cheap'' classical communication lines it is often the case that tasks that are classically difficult or impossible to achieve can in fact be accomplished. 

Such a context is broad enough to encompass quite a vast range of scenarios, formalized and explored in these years: from the nowadays mature technology of quantum cryptography~\cite{gisin} to one-way/measurement-based quantum information processing~\cite{browne} and distributed quantum computing. In a nutshell, these last two examples see their success in the processing of quantum information via a clever use of quantum teleportation 
~\cite{bennett,pan}. In fact, this can be thought as the fundational building-block for distributed quantum information processing (QIP). 
Although an exhaustive account is beyond the scope of the present paper, it has to be mentioned that a whole wealth of studies has followed the original protocol for teleportation by Bennett {\it et al.}~\cite{kim,yeo,li,ge,7,8,9,10,11}. In particular, motivated by the research for a better understanding of the still elusive trade-off existing between purity and entanglement, the performance of teleportation via mixed entangled states has been studied (notably in~\cite{darianobowen}). This has revealed that, differently from the standard teleportation protocol via a maximally-entangled pure state, the use of a mixed-state resource corresponds to a noisy channel of some sort for the state to be teleported. These considerations have been extended to the case where the state of a composite system or, more specifically, its entanglement content, is the focus of our attention. In particular, Werner states under the use of {\it negativity} as an entanglement measure~\cite{pereshorodecki} have been shown to be potential channels for such an {\it entanglement teleportation} scheme~\cite{kim}. It is interesting to go beyond such results and fully understand what one has to expect, in terms of teleportation capabilities, when a non-ideal state is used and which are the implications, in quantitative terms, of the use of different characterization tools. While the first point is made sensible by considering that ever-present environmental mechanisms always conspire to spoil an ideal resource, the second highlights the ambiguities that might well be in order when mixed entangled states are involved. The progress of distributed QIP towards physical implementability has to pass through the assessment of both such points, in a broad context. 

Given that in a real experiment we will have to consider mixed states as our primary resource, it is quite meaningful to use those presenting the most promising features. Here we thus consider the case where maximally entangled mixed states (MEMS) ~\cite{munro}, which maximize the available degree of entanglement of a mixed state per assigned global mixedness, embody the required channel in an entanglement teleportation protocol. We adopt concurrence~\cite{wooters} and linear entropy~\cite{bose} as our figures of merit for MEMS parameterization. We characterize the performances of such boundary states as quantum channels for entanglement teleportation, spotting out a significant dependence of the scheme's efficiency on the {\it form} of the entangled state to teleport and an even more intriguing sensitivity to the specific choice of entanglement measure used in order to quantify the teleported entanglement. In consideration of the very recent success in experimentally generating and characterizing MEMS~\cite{cinelli,peters} and the necessity of a structured program for the assessment of their usefulness for coherent QIP (of which our study can rightly be regarded as one of the first significant steps), we also propose a readily-implementable linear-optics set-up able to test the quantitative results put forward in our investigation.

The remainder of the paper is organized as follows. Section~\ref{MEMS} introduces the main aspects of MEMS and discusses some of the quantitative tools used  in our study. Section~\ref{analysis} reports our main results and highlights the strengths and limitations of using these states as entanglement teleportation channels. In Sec~\ref{exp} we propose a linear-optics set-up for the test of our theoretical assessment. In Section~\ref{conclusion} we give some concluding remarks.

\section{Maximally Entangled Mixed States and Entanglement Teleportation}
\label{MEMS}
The class of MEMS has been shown to be of particular interest. These bipartite states are endowed with the maximum allowed amount of entanglement a state can possess when its global mixedness has been assigned~\cite{munro}. With this in mind, one often finds it insightful, although not quite rigorous, to view them as mixed-state versions of pure Bell states~\cite{nielson}. We believe the potential of these states as QIP resources is yet to be satisfactorily explored and our study can be seen as one of the first steps along these lines~\cite{commentIndia}.

\begin{figure}[b]
{\bf (a)}\hskip3cm{\bf (b)}
\psfig{figure=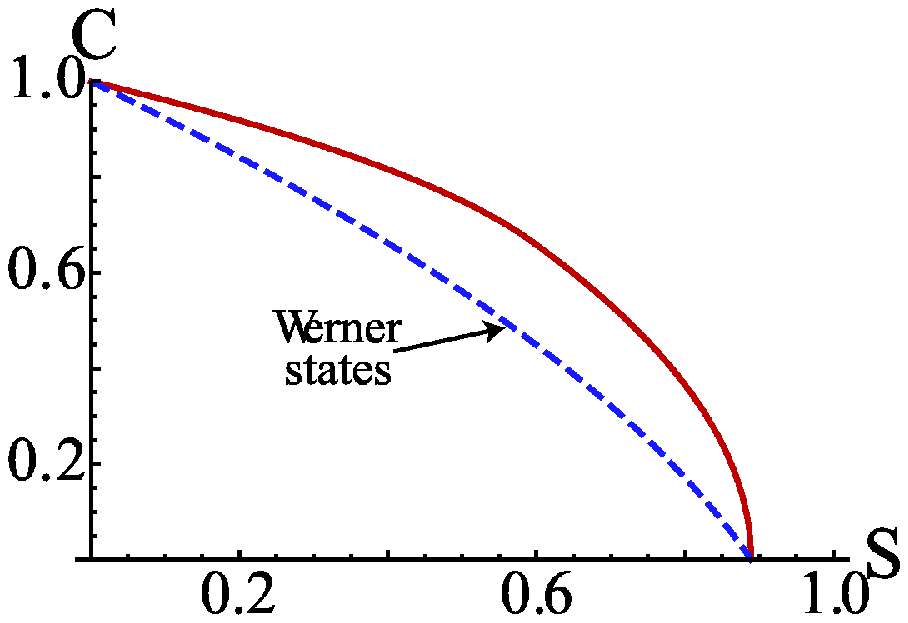,width=4.3cm,height=3.2cm}~\psfig{figure=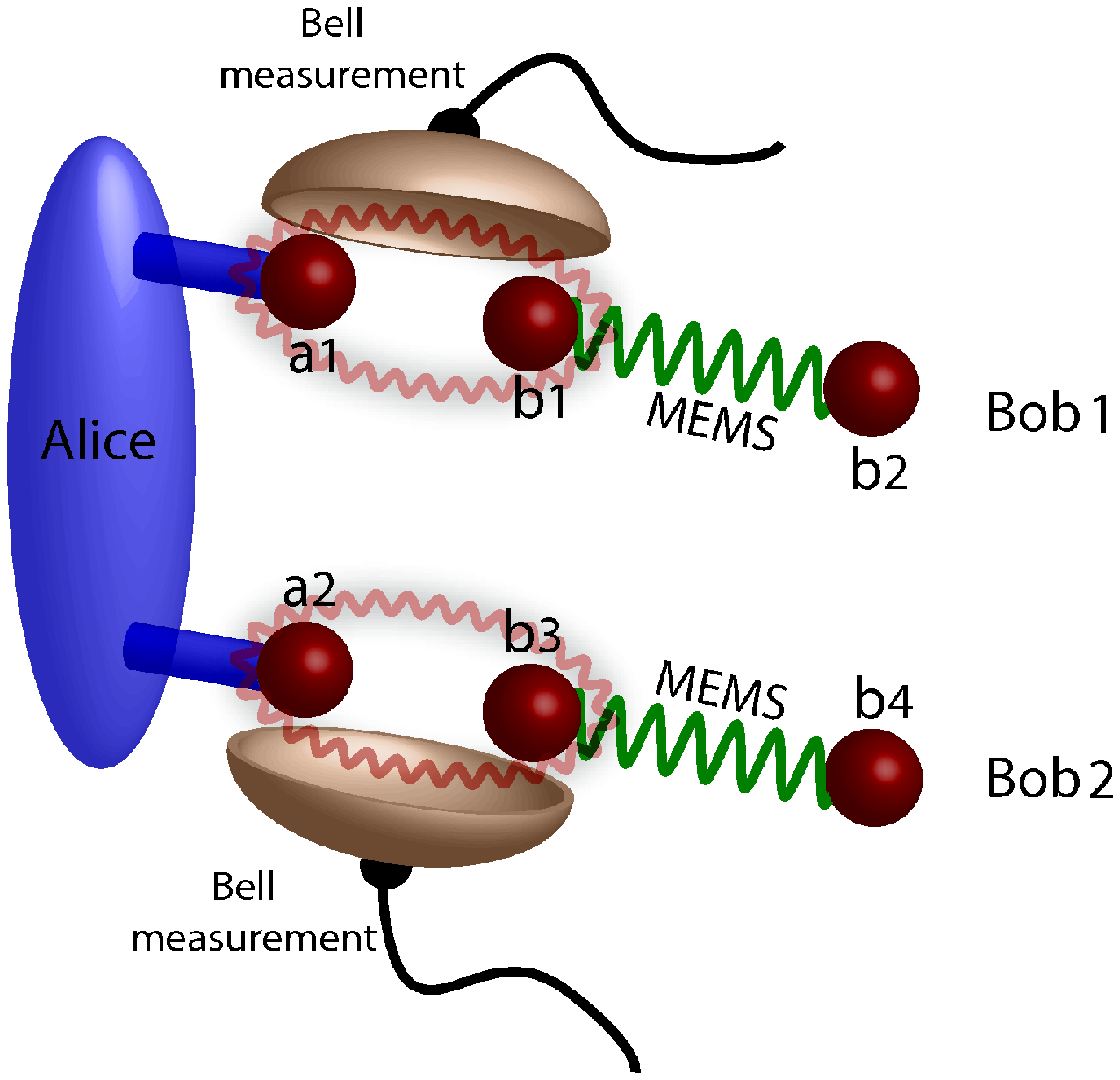,width=4.5cm,height=3.7cm}
\caption{(Color Online) {\bf (a)} MEMS boundary in the concurrence $C$ versus linear entropy $S$ plane. For comparision, we also plot the two qubit Werner states. {\bf (b)} Schematic representation of the considered teleportation protocol. Alice is in possession of the unknown state, and each Bob sends her one qubit out of their respective MEMS pair. Alice then performs Bell-state measurements on the qubits she has and classically communicates to each Bob which outcome has been obtained. Upon receipt of this information, a referee checks the degree of entanglement shared by the qubits held by Bob's.}
\label{scheme}
\end{figure}

In dealing with mixed entangled states, one immediately faces ordering problems arising from the ambiguties associated with different bipartite entanglement measures~\cite{virmani}. Despite physically representing the same boundary, the parameterization for MEMS changes depending on the entanglement measures being used~\cite{munro}. As anticipated, here we consider the parameterization of boundary states corresponding to the concurrence $C$ of a state $\rho$~\cite{wooters}, which is defined as
\begin{equation}
C= \max [0,\sqrt{\lambda_1}-\sum^4_{j\ge{2}}\sqrt{\lambda_j}].
\end{equation}
Here $\lambda_1\ge\lambda_j~(j=2,3,4)$ are the eigenvalues of $\rho(\hat{\sigma}_2\otimes\hat{\sigma}_2) \rho^* (\hat{\sigma}_2\otimes\hat{\sigma}_2)$, we have used the ordering $\hat{\sigma}_1=\hat{\sigma}_x,\hat{\sigma}_2=\hat{\sigma}_y,\hat{\sigma}_3=\hat{\sigma}_z,\hat{\sigma}_4=\openone$ and $\hat{\sigma}_{x,y,z}$ are the three Pauli spin operators. As a convenient measure for the mixedness of a state, we use linear entropy~\cite{bose}
\begin{equation}
S=\frac{4}{3}(1-\text{Tr}[(\rho)^2]).
\end{equation}
The corresponding one-parameter class of MEMS comprises two families of density matrices, as shown in~\cite{munro}, which read 
\begin{equation}
\label{mems}
\rho^1 = 
\left(
\begin{array}{llll}
 \frac{r}{2} & 0 & 0 & \frac{r}{2} \\
 0 & 1-r & 0 & 0 \\
 0 & 0 & 0 & 0 \\
 \frac{r}{2} & 0 & 0 & \frac{r}{2}
\end{array}
\right),~~~\rho^2 =
\left(
\begin{array}{llll}
 \frac{1}{3} & 0 & 0 & \frac{r}{2} \\
 0 & \frac{1}{3} & 0 & 0 \\
 0 & 0 & 0 & 0 \\
 \frac{r}{2} & 0 & 0 & \frac{1}{3}
\end{array}
\right),
\end{equation}
where $\rho^1$ ($\rho^2$) holds for $r\in[2/3,1]$ ($r\in[0,2/3]$). Entanglement teleportation via Werner-state channels has been shown in~\cite{kim}. 
Interestingly, under the suitable entanglement measure of negativity~\cite{pereshorodecki}, Werner states {\it belong} to MEMS. This is not the case if concurrence is used instead, as illustrated in  Fig.~\ref{scheme} {\bf (a)}. Such an apparently innocent difference brings about, in reality, a series of significant differences, which will be discussed later on. 

It is useful, at this stage, to briefly discuss the basic features of the entanglement teleportation protocol, whose steps will be formalized in the next Section. The protocol is schematically shown in Fig.~\ref{scheme} {\bf(b)} and can be conveniently thought of as a three-player quantum game. Alice, who is in possession of two qubits (labelled $a_1$ and $a_2$) prepared in an unknown state, wishes to teleport their entanglement to two distant, initially uncorrelated qubits. She is aided in her task by two other agents, Bob1 and Bob2, each managing a two-qubit MEMS state belonging to either family $\rho^1$ or $\rho^2$. Bob1 (Bob2) holds qubits $b_1\&b_2$ ($b_3\&b_4$). They send Alice one qubit of their pairs, say $b_1$ and $b_3$ respectively. Alice then performs joint Bell-state measurements (BSM's) on the qubits on the pairs $a_1\&b_1$, $a_2\&b_3$ and transmits four bits of classical information (on the outcome of her measurements) to each Bob. A referee then checks the degree of teleported entanglement. Had the channels been pure, upon application of proper unitaries on qubits $b_2$ and $b_4$, Bob1 and Bob2 would have recreated the input state Alice had at the start. We will see that, through MEMS channels, they can share some (even quite considerable) fraction of the degree of entanglement that was to be teleported, although the success of the protocol crucially depends on the choice of MEMS to use and the form of the input state.

\section{MEMS as Entanglement Teleportation Channels}
\label{analysis}
In this Section we present our main analysis. We start outlining the formal approach to entanglement teleportation via MEMS. We then address our central results, which cover three distinct scenarios. First we present how the channels perform when attempting to teleport pure target states as parameterized by their input entanglement. This will be sufficient to gain some fundamental insight into the efficiency of MEMS as teleportation channels and expose the sensitivity of the scheme to the form of the target state. Next, we examine the state fidelity between target and teleported states, finding unexpected differences with respect to previous entanglement teleportation protocols based on Werner states and the use of negativity~\cite{kim}. Finally, the analysis is extended to arbitrary (in general) mixed target states with the use of some numerical tools.
 
Let us start by providing the general framework for our quantitative study. Any arbitrary (pure or mixed) two-qubit state can be written as~\cite{nielson}
\begin{equation}
\label{general}
\begin{split}
\varrho_{a_1,a_2}&=\frac{1}{4}[\openone^{a_1}\otimes\openone^{a_2}\!+\!\sum^3_{k=1}(\beta_{k}\sigma^{a_1}_{k} \otimes \openone^{a_2}+\gamma_{k}\openone^{a_1}\otimes\sigma^{a_2}_{k})\\
&
+\sum^3_{k,l=1}\!\chi_{kl}\sigma^{a_1}_{k} \otimes \sigma^{a_2}_{l}],
\end{split}
\end{equation}
where ${\bm \beta}$ and ${\bm \gamma}$ are the Bloch vectors of qubit $a_1$ and $a_2$ respectively~\cite{nielson}, while the matrix ${\bm \chi}$ encompasses their correlations. The composite state of the input target state and two MEMS channels is thus expressed as
\begin{equation}
\rho=\varrho_{a_1,a_2}\otimes\rho^{i}_{b_1,b_2}\otimes\rho^{k}_{b_3,b_4},
\end{equation}
where $i,k=1,2$. The teleportation is then performed, as qualitatively described in Sec.~\ref{MEMS}, by implementing two BSM's. These are formally described by the use of state projectors $\hat{\bm \Pi}^\mu_{a_1,b_1}\otimes\hat{\bm \Pi}^\nu_{a_2,b_3}~(\mu,\nu=1,..,4)$ with, for instance, $\hat{\bm \Pi}^\mu_{a_1,b_1}=|{\cal B}_\mu\rangle_{a_1,b_1}\langle{\cal B}_{\mu}|$ and $\ket{\cal B_\mu}_{a_1,b_1}\!=\!\hat{\sigma}^{a_1}_\mu\!\otimes\!\openone^{b_1}\ket{\Psi^+}_{a_1,b_1}$, where we have introduced the elements of the Bell-state basis $\ket{\Phi^{\pm}}=(\ket{00}\pm\ket{11})/\sqrt{2}$, $\ket{\Psi^{\pm}}=(\ket{01}\pm\ket{10})/\sqrt{2}$. {\it In what follows, motivated by the specific experimental set-up we propose later, for which full Bell-state discrimination is impossible, we assume a ``rigid'' projection always performed onto $\ket{\Phi^+}_{a_1,b_1}\ket{\Phi^+}_{a_2,b_3}$}, so that we shall be interested in quantifying the amount of {\it teleported} entanglement of the output state
\begin{equation}
\label{formal}
\rho^{out}_{b_2,b_4}={\cal N}\text{Tr}_{a_1,a_2,b_1,b_3}[(\hat{\bm \Pi}^1_{a_1,b_1}\otimes\hat{\bm \Pi}^1_{a_2,b_3})\rho],
\end{equation}
retrieved by Bob1 and Bob2 after Alice's measurements (here ${\cal N}$ is the normalization factor). Beside any experiment-related argument, it is worth noticing that for rigid projections onto $\ket{\Phi^+}$,  if the channels used by Bob's is the tensor product of two $\rho^1$ states, for $r=1$ ({\it i.e.} in the ideal-channel case) no by-product local unitary is required on qubits $b_2$ and $b_4$ in order to retrieve the input target state.  

\subsection{Teleportation of pure-state entanglement via MEMS}
\begin{figure}[b]
\psfig{figure=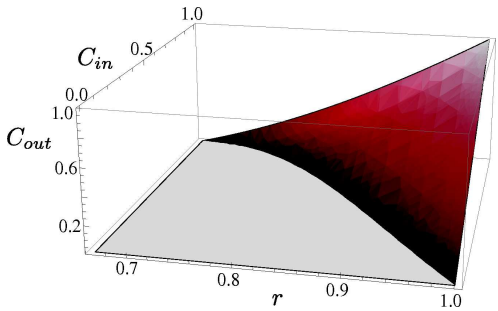,width=6.5cm,height=4.cm}
\caption{(Color online) Teleported entanglement $C_{out}$ plotted against channel quality $r$ and input concurrence $C_{in}$ for a pure target state as defined in Eq.~(\ref{purephi}). We assume the use of $\rho^{1}_{b_1,b_2}\otimes\rho^{1}_{b3,b4}$ as the teleportation channel. The ability to teleport entanglement is dependent on both the quality of the channels in use and the degree of quantum correlations of the target state. The flat region clearly indicates the fact that there is a `lower bound' to the target entanglement for set quality $r$ of the channel. Any value of concurrence $C_{in}$ below this threshold yields a separable teleported state.}
\label{MEMS1pure}
\end{figure}

We start addressing the performances of MEMS as channels for entanglement teleportation by considering the significant case of a pure input state such as ($\alpha\in\mathbb{R}$)
\begin{equation}
\label{purephi}
\ket{\phi}_{a_1,a_2}=\alpha \ket{00}_{a_1,a_2} + \sqrt{1-\alpha^{2}} \ket{11}_{a_1,a_2}.
\end{equation}
The assumptions underlying our approach is that the players of the game {do not know} the degree of entanglement within the state Alice is provided, although they are aware of the general form of it.  
We now incorporate the amount of entanglement to teleport by parameterizing Eq.~(\ref{purephi}) in terms of its concurrence $C_{in}$ as
\begin{equation}
\alpha=\sqrt{\frac{1+\sqrt{1-C_{in}^2}}{2}}.
\end{equation}
Let us start by taking the two channels as belonging to the large-entanglement family $\rho^{1}$, both having the same quality ({\it i.e.} they have the same $r$). It is then a relatively straightforward calculation to determine how the teleported entanglement depends on the quality of the channels and the amount of initial entanglement. Although an explicit analytic formula for the teleported concurrence can indeed be found, its cumbersome and lengthy nature makes it unsuitable to be reported here. The important features are nontheless visible in Fig.~\ref{MEMS1pure}, where we plot the teleported entanglement $C_{out}$ against the input concurrence $C_{in}$ and the quality of the channel $r$: the higher the quality of the channel, the more entanglement we see successfully teleported. Ideally, for $r=1$, the output entanglement is equal to $C_{in}$. More interestingly, though, there is a cut-off in the quality of the channel below which no entanglement can be teleported (any entangled target state is mapped onto a separable one). This region is clearly shown by the flat area in Fig.~\ref{MEMS1pure}. The threshold value of $r$ beyond which one is guaranteed that any entangled target state is mapped into an entangled teleported one (although of a different form than the target) can be easily found by looking at the value of $r$ at which $C_{out}=0$ for a set $C_{in}$. This gives
\begin{equation}
r_{t}=\frac{4-2C_{in}+4\sqrt{1-C^2_{in}}}{3+5\sqrt{1-C^2_{in}}}.
\end{equation}
For an assigned value of $C_{in}$, any $r>r_t$ is associated with $C_{out}>0$, thus signaling success of the teleportation protocol.

The teleportation capabilities of the scheme are rather different if we take the channels as belonging to the lower-entanglement MEMS class encompassed by $\rho^{2}$. It turns out, indeed, that this type of channel is unable to teleport the entanglement carried by a state like Eq.~(\ref{purephi}), regardless of the amount of input concurrence being present. This result contrasts strikingly with previous findings. We can gain insight into why this should be the case by considering the point at which both families of MEMS defined in Sec~\ref{MEMS} overlap, {\it i.e.} $r=2/3$. From Fig.~\ref{MEMS1pure} when the channel is of its lowest quality ($r=2/3$) no entanglement can be teleported and it is thus quite intuitive that, by further reducing the quality of the channel, the ability to teleport entanglement will decrease as well.

\begin{figure}[t]
{\bf (a)}\hskip4cm{\bf (b)}
\psfig{figure=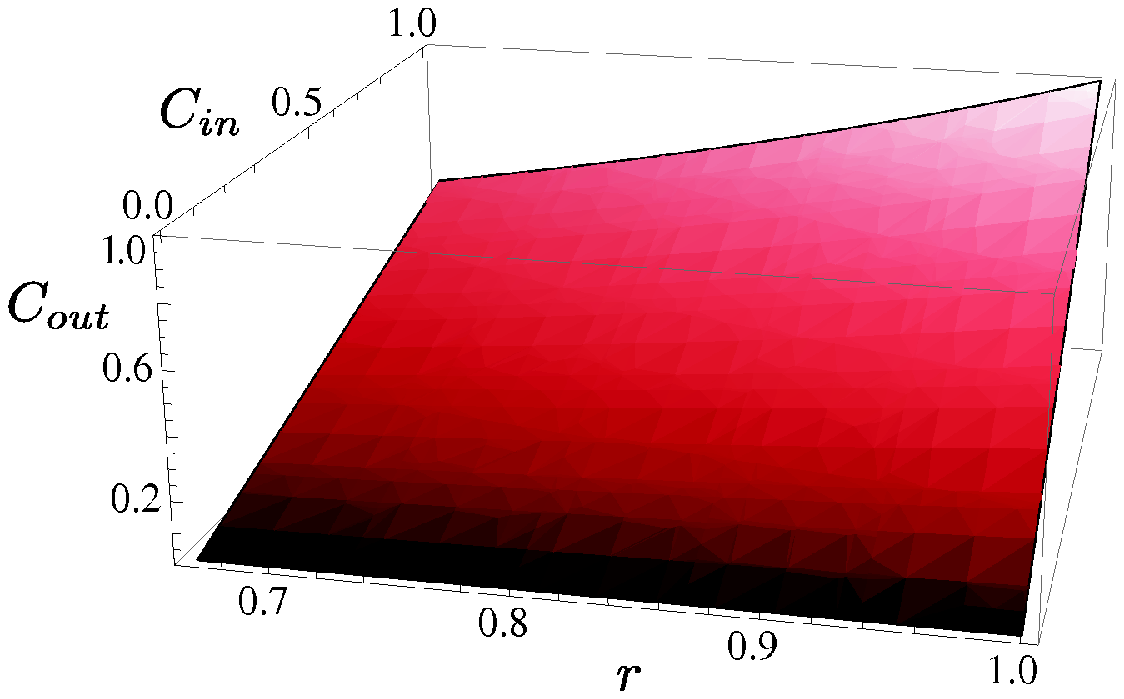,width=4.0cm,height=3.2cm}~~\psfig{figure=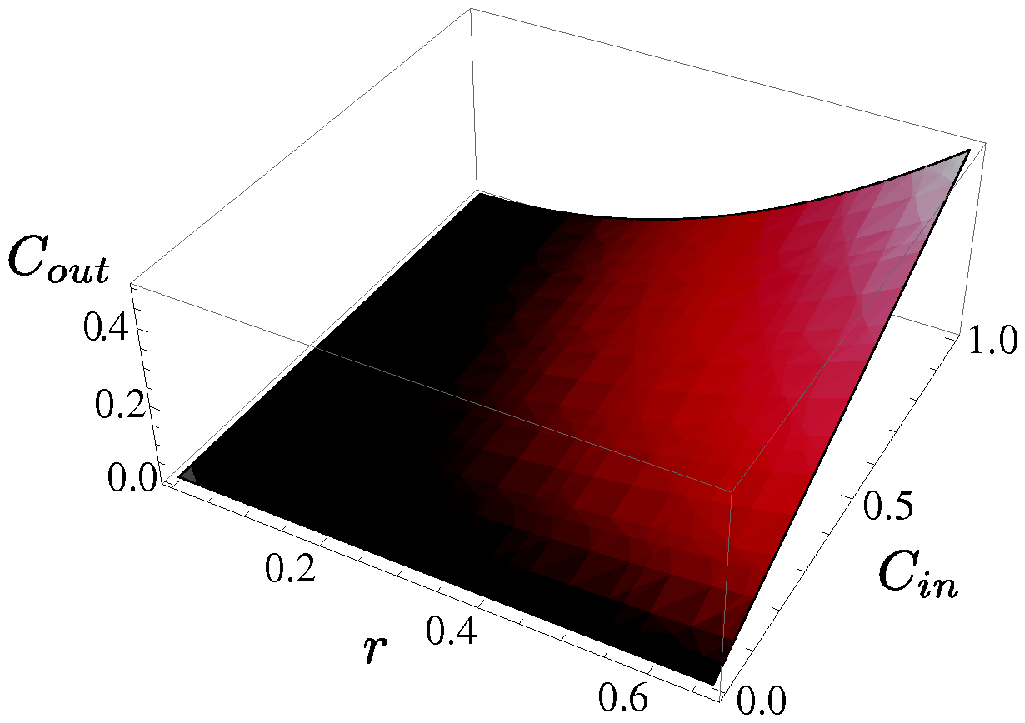,width=4.0cm,height=3.2cm}
\caption{(Color Online) {\bf (a)} Teleported entanglement $C_{out}$ against target-state concurrence $C_{in}$ and channel quality $r$ for teleportation via the MEMS channel $\rho^{1}_{b_1,b_2}\otimes\rho^{1}_{b_3,b_4}$ when the target state is of the form given by Eq.~(\ref{purepsi}). {\bf (b)} Same as in panel {\bf (a)} but using $\rho^{2}_{b_1,b_2}\otimes\rho^{2}_{b_3,b_4}$.}
\label{psitypeorig}
\end{figure}

The situation is notably different when we consider an input class of entangled states that is locally equivalent to Eq.~(\ref{purephi}), such as
\begin{equation}
\label{purepsi}
\ket{\psi}_{a_1,a_2}=\alpha \ket{01}_{a_1,a_2} + \sqrt{1-\alpha^{2}} \ket{10}_{a_1,a_2}.
\end{equation}
Obviously, $\ket{\phi}_{a_1,a_2}$ and $\ket{\psi}_{a_1,a_2}$ are equally entangled. And yet, a significant difference in the performance of the teleportation channels is found, as shown in Figs.~\ref{psitypeorig}. By taking both channels of the form $\rho^{1}$, as for the case shown in Fig.~\ref{psitypeorig} {\bf (a)},  the ability to teleport entanglement is again dependent on $r$ and the amount of input entanglement. Differently from Eq.~(\ref{purephi}), regardless of the quality of the channel or the amount of input entanglement, this situation corresponds to an always successful teleportation, although the input concurrence can be strongly degraded by the process itself. This time, a  concise analytic formula is possible
\begin{equation}
\label{primo}
C_{out}=\frac{r}{2-r}C_{in},
~~~~~~(r\in [2/3,1]).
\end{equation}
The differences with respect to the case associated with state $\ket{\phi}$ extend to the use of $\rho^{2}$, as shown in Fig~\ref{psitypeorig} {\bf (b)}, where we now find that entanglement is faithfully teleported. Again, the corresponding analytic expression for $C_{out}$ is quite informative and reads
\begin{equation}
\label{secondo}
C_{out}=\frac{9r^2}{8}C_{in}, 
~~~~~~(r\in [0,2/3]).
\end{equation}
The quadratic functional behavior of the above equation against $r$ explains the evident sensitivity of this instance of the protocol to the channels' properties and, in general, the poorer performance of the teleportation scheme. The difference between Eq.~(\ref{primo}) and~(\ref{secondo}) arise from the radically different dependence of $\rho^1$ and$\rho^2$ on $r$. The latter having fixed populations (cfr. Eq.~(\ref{mems})), gives rise to a more fragile output concurrence.

The reasons behind the differences in performance associated with input target states having different forms can be explicitly related to the sort of effective {\it noisy channels} embodied by the teleportation protocol when $\rho^1$ or $\rho^2$ are chosen. This observation can be cast in more quantitative terms by considering the following approach. We start rewriting Eqs.~(\ref{mems}) in the basis of the {output} qubits ({\it i.e.} $b_2$ and $b_4$) as
\begin{equation}
\begin{aligned}
\rho^j_{b_kb_{k+1}}&=\hat{A}^j_{b_k}\otimes\miniproj{0}{b_{k+1}}{0}+\hat{B}^j_{b_k}\otimes\miniproj{0}{b_{k+1}}{1}\\
&+\hat{C}^j_{b_k}\otimes\miniproj{1}{b_{k+1}}{0}+\hat{D}^j_{b_k}\otimes\miniproj{1}{b_{k+1}}{1}
\end{aligned}
\end{equation}
with $j=1,2$, $k=1,3$ and $\hat{\cal O}^j_{b_k}=\{\hat{A}^j_{b_k},\hat{B}^j_{b_k},\hat{C}^j_{b_k},\hat{D}^j_{b_k}\}$ a set of single-qubit operators defined in qubit $b_k$'s Hilbert space as shown in Ref.~\cite{comment}. With this decomposition it is straightforward to see that the effects of the teleportation protocol on the logical input state whose entanglement we are interested in can all be gathered by evaluating the 16 expectation values $_{a_1,b_1}\minisand{\Phi^+}{[\miniproj{l}{a_1}{m}\otimes{(\hat{\cal O}^j_{b_1})_{p}}]}{\Phi^+}_{a_1,b_1}$ and $_{a_2,b_3}\minisand{\Phi^+}{[\miniproj{l}{a_2}{m}\otimes{(\hat{\cal O}^j_{b_3})_{p}}]}{\Phi^+}_{a_2,b,3}$, where $l,m=0,1$ and $p=1,..,4$. It turns out that many of them are identically zero, which simplifies the calculation considerably and leaves us with the explicit form of the logical output two-qubit state encoded into qubits $b_2\&b_4$. From this, the action of the effective noisy mechanism can be inferred. For the case of $\rho^2$ channels and $\varrho_{a_1,a_2}=\miniproj{\phi}{a_1,a_2}{\phi}$, we obtain
\begin{equation}
\label{rho2Phi}
\rho^{out}_{b_2,b_4}=
\begin{pmatrix}
\frac{\alpha^2}{1+3\alpha^2}
&0&0&\frac{9r^2\alpha\sqrt{1-\alpha^2}}{4(1+3\alpha^2)}\\
0&\frac{\alpha^2}{1+3\alpha^2}&0&0\\
0&0&\frac{\alpha^2}{1+3\alpha^2}&0\\
\frac{9r^2\alpha\sqrt{1-\alpha^2}}{4(1+3\alpha^2)}&0&0&\frac{1}{1+3\alpha^2}
\end{pmatrix},
\end{equation}
whose partial transposition never gives rise to a negative eigenvalue for the parameterization of $\ket{\phi}_{a_1,a_2}$ used in this work. This implies that the output state is always separable in virtue of the Peres-Horodecki criterion~\cite{PPT}. On the other hand, if it is $\ket{\psi}$-type entanglement that has to be teleported, through the same method we get
\begin{equation}
\label{rho2Psi}
\rho^{out}_{b_2,b_4}=
\begin{pmatrix}
0&0&0&0\\
0&\frac{\alpha^2}{2}&\frac{9r^2\alpha\sqrt{1-\alpha^2}}{8}&0\\
0&\frac{9r^2\alpha\sqrt{1-\alpha^2}}{8}&\frac{1-\alpha^2}{2}&0\\
0&0&0&\frac{1}{2}
\end{pmatrix},
\end{equation}
whose partial transposed form is always non-positive, as far as $C_{in}\neq{0}$. Analogous analyses can be performed with respect to the choice of $\rho^1$ as a channel, getting results in perfect agreement with our quantitative study of the output concurrence. One can thus grasp an intuition of the effects of the MEMS-affected teleportation protocol: a $\rho^2$-based channel originates a logical noisy mechanism where the even-parity subspace ({\it i.e.} the one spanned by the logical $\ket{00}$ and $\ket{11}$) is depleted more severely than what happens to the odd-parity one. The coherences in each logical state are reduced in magnitude by $r^2$, as it is intuitively understood considering the double-channel nature of entanglement teleportation. In effect, an input state having a large projection onto $\ket{\Psi^\pm}$ would see its entanglement surviving the teleportation process, differently from what happens to the more {\it exposed} case of states close to $\ket{\Phi^\pm}$. We are currently investigating the relations between this effect and the known resilience of $\rho^2$ type of MEMS to certain types of collective noise~\cite{campbell}.

\begin{figure}[b]
\psfig{figure=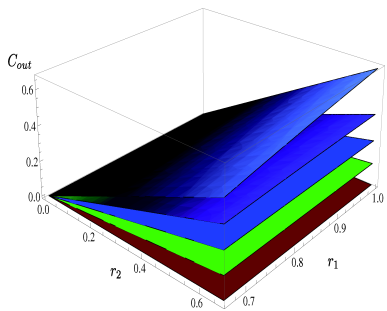,width=5cm,height=4cm}
\caption{(Color online) Teleported entanglement plotted against channel quality for channels belonging different families of MEMS. $r_{j}$ denotes the quality of the channel defined by $\rho^{j}~(j=1,2)$. Each curve corresponds to an increasing amount of target concurrence $C_{in}$ from 0 to 1 in steps of 0.25.}
\label{both}
\end{figure}

For completeness, the case where two channels belong to different MEMS families, such as $\rho^{1}_{b_1,b_2}\otimes\rho^{2}_{b_3,b_4}$, has also been considered. We have found a quantitative compromise between the extreme situations addressed so far: the teleported entanglement is dependent on the quality of both channels and, as with the previous analysis, the higher the target concurrence, the more entanglement we see teleported. The results corresponding to the case of a target state of the form of Eq.~(\ref{purephi}) are shown in Fig.~\ref{both}.

\subsection{Teleportation fidelity and target entanglement}
\begin{figure}[t]
{\bf (a)}\hskip3cm{\bf (b)}
\psfig{figure=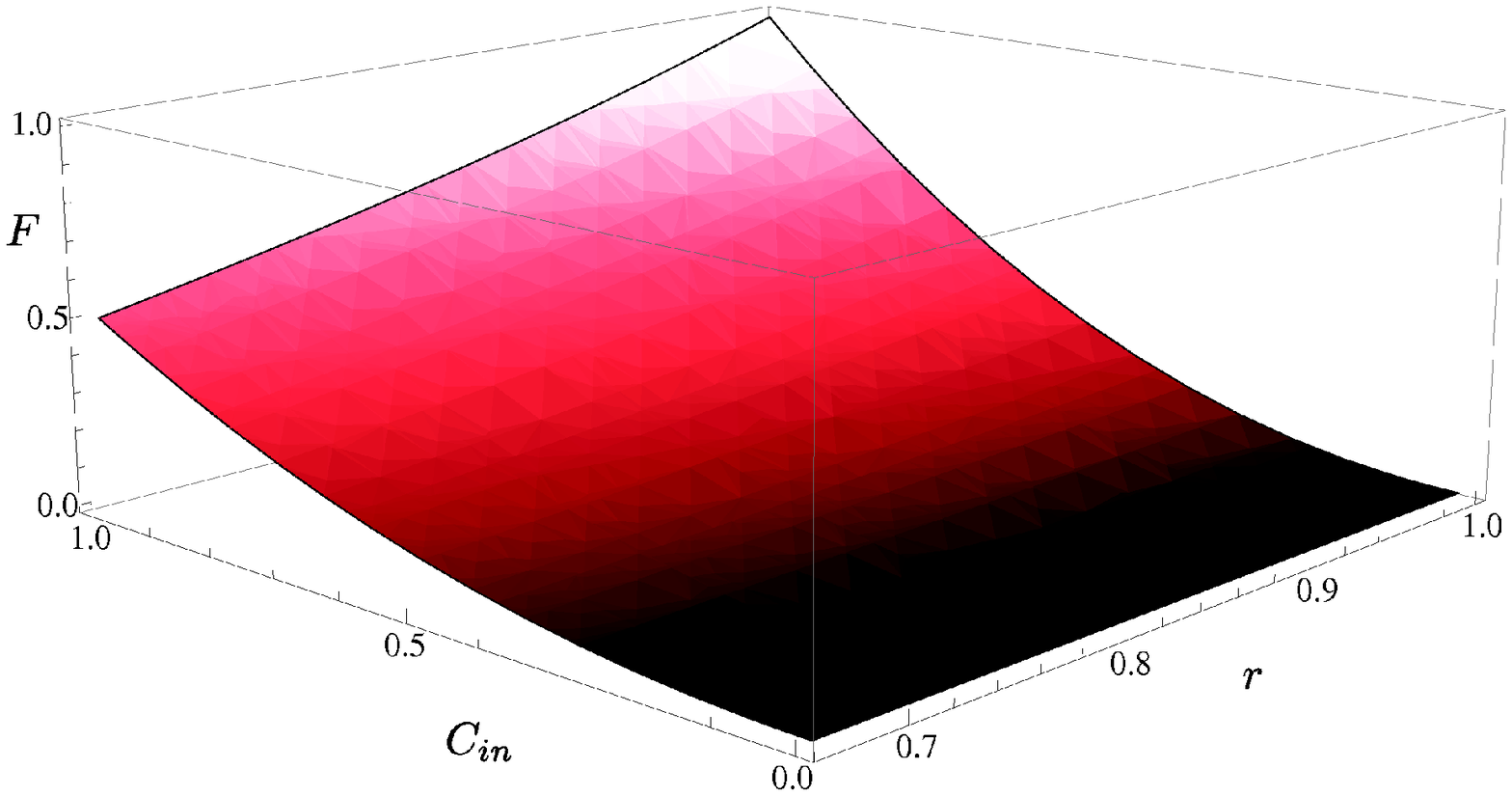,width=4.0cm,height=3.2cm}~~\psfig{figure=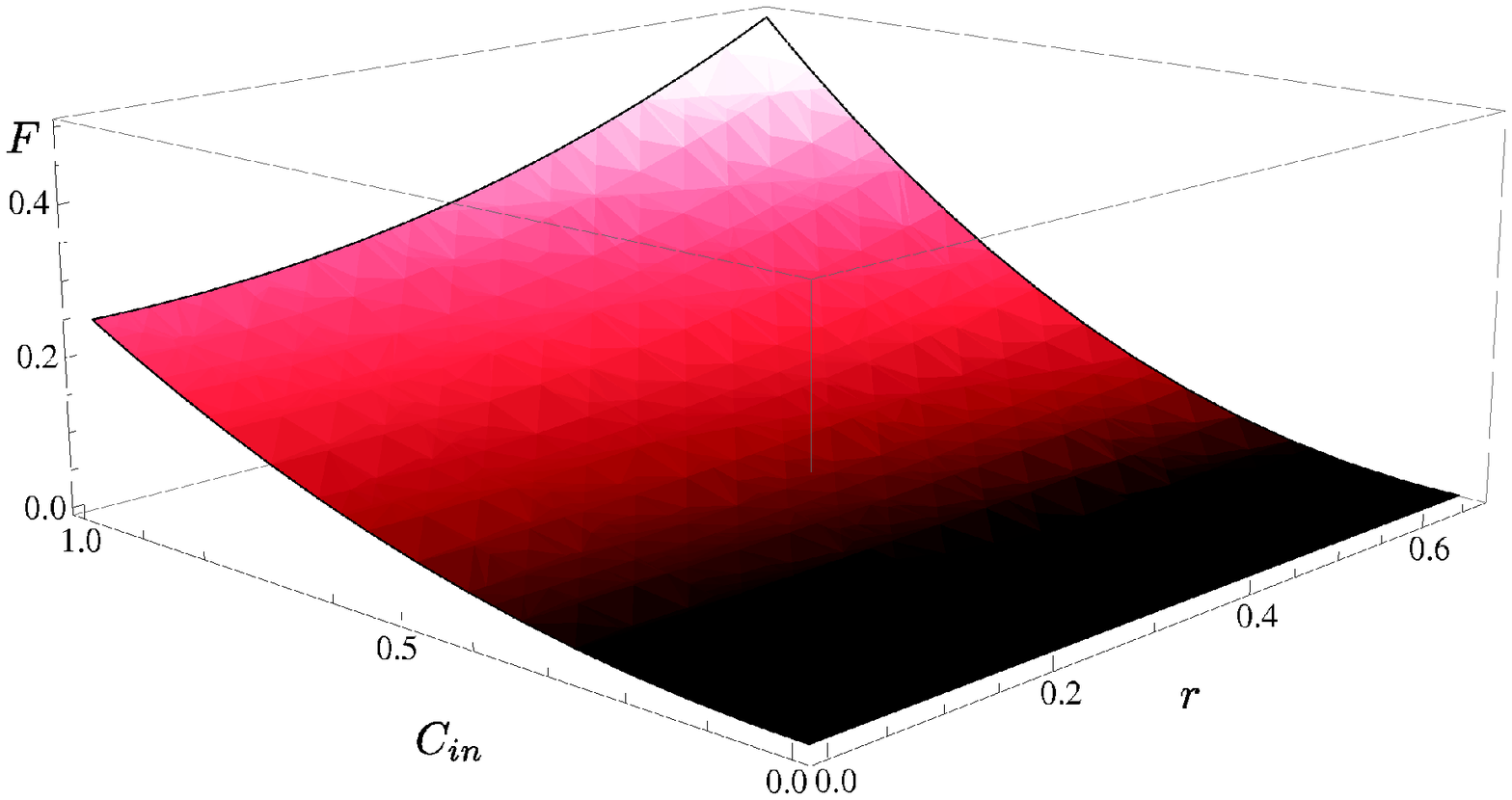,width=4.0cm,height=3.2cm}
\caption{(Color online) {\bf (a)} State fidelity of the teleported state with respect to the initial target state, when both channels belong to $\rho^{1}$. We note the increase in fidelity as the input entanglement is increased.  {\bf (b)} As for panel {\bf (a)} but both channels belonging to $\rho^{2}$.}
\label{fidelity}
\end{figure}

Until now we have been concerned only with the degree of entanglement being teleported by means of the chosen form of channels. However, one might well wonder how similar the teleported state is to the original target state. We determine the closeness of two states using quantum state fidelity~\cite{nielson}. Its definition, adapted to the aims of our work, is given by
\begin{equation}
F=\langle\text{target}|\rho^{out}_{b_2,b_4}|\text{target}\rangle
\end{equation}
with $\ket{\text{target}}$ the target input state.
It is already a well known result that teleportation via local operations and classical communication (LOCC) can at best yield a fidelity of $F=2/3$. If we consider again the pure state in Eq.~(\ref{purepsi}), we can easily calculate the fidelity of the teleported state depending on the quality of the channel and the input entanglement. Abandoning for a while the rigidity of our BSM's, it is obvious that the actual value of $F$ will be very sensitive to the outcomes of the BSM's performed at Alice's site. Consequently, depending on the results of the projections being performed, the appropriate local by-product operation on the teleported state is applied in a way so as to regain the highest fidelity. The results are summarized in Fig.~\ref{fidelity} and clearly show we can do better than LOCC provided we are dealing with the highly entangled channels $\rho^{1}$. We also find an interesting difference with respect to previous entanglement teleportation studies: regardless of the quality of the channel, state  fidelity always increases with the amount of input entanglement. This is contrary to what was found in Ref.~\cite{kim}, where the teleportation protocol performed via Werner states gives rise to a state fidelity that decreases when $C_{in}$ is increased. It is straightforward to determine elegant analytic expressions for the fidelity in terms of the channel quality and input entanglement, reading
\begin{equation}
\begin{aligned}
F_1&=\frac{9r^2+4}{16}C_{in}^{2}~~~~~ 
(r \in[0,2/3]),\\
F_2&=\frac{r}{2-r}C_{in}^2~~~~~
(r\in [2/3,1]),
\end{aligned}
\end{equation}
where $F_{j}$ is the state fidelity obtained upon choice of $\rho^j_{b_1,b_2}\otimes\rho^j_{b_3,b_4}$.
This result leads us to an interesting conclusion. As we have already commented, Werner states correspond exactly to MEMS if entanglement (mixedness) is quantified through negativity (linear entropy) and, by studying their performance as channels for the teleportation of entanglement~\cite{kim}, it has been seen that entanglement is fragile. Yet, by changing picture and looking at a different entanglement measure, we have a different parameterization for MEMS and, quite unexpectedly, there is a very different performance when the teleported state fidelity is considered: concurrence shows some robustness. This leads us to conclude that the efficiency of MEMS for teleportation is sensitive to the entanglement measure being taken.

\subsection{Teleportation of arbitrary-state entanglement}
By now, we have exhaustively studied the performance of MEMS in teleporting pure-state entanglement as quantified by their concurrence. A natural progression is to see how the channels perform when the initial state is mixed and we relax the assumption on knowing the ``form" (but not the degree of entanglement) of the input state. This move necessitates us to depart from the analytic studies at the root of our analysis so far and rely on numerics to quantitively examine the performance of the scheme. Let us start again with our target state as expressed in its most general form of Eq~(\ref{general}). Our approach is then based on the construction of a large sample of random density matrices. Each element of the sample is then equated to  Eq.~(4) in order to determine the corresponding ${\bm \beta}$, ${\bm \gamma}$ and ${\bm \chi}$. The protocol is then ran as before.

\begin{figure}[t]
\centerline{{\bf (a)}\hskip3.5cm{\bf (b)}}
\psfig{figure=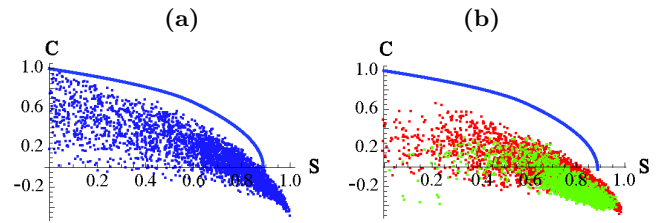,width=8.5cm,height=2.5cm}
\caption{(Color Online) {\bf (a)}: We show the position in the $C-S$ plane occupied by 3000 two-qubit mixed density matrices, randomly generated in a uniform way according to Haar measure~\cite{nielson}. The MEMS boundary (full line) is also given as a reference. {\bf (b)}: Each of the points shown in panel {\bf (a)} is mapped into the corresponding one in the  output concurrence versus purity plane. The (red) dark points correspond to the use of $\rho^1_{b_1,b_2}\otimes\rho^1_{b3,b4}$ while the lighter (green) ones correspond to the channel $\rho^2_{b_1,b_2}\otimes\rho^2_{b_3,b_4}$. In both panels, in order to show conservaton of the number of points, we plot $\sqrt{\lambda_1}-\sum^4_{j\ge{2}}\sqrt{\lambda_j}$ for the concurrence.}
\label{mixd}
\end{figure}

The algorithm used in order to generate the sample of random density matrices exploits the method described in Ref.~\cite{random}. It passes through the generation of an arbitrary unitary operator $\hat{U}$ (valid in principle for any number of qubits) parameterized in terms of a proper number of  Eulers angles, each being randomly determined from a uniformly distributed set of values. Such a unitary, which is in general entangling, is then applied to a random normalized diagonal matrix $D$. We thus construct our (in general) mixed entangled target state as
\begin{equation}
\varrho=\hat{U} D \hat{U}^{\dagger}.
\end{equation}
Fig.~\ref{mixd} {\bf (a)} shows the position occupied by 3000 of such random states in the $C-S$ plane. On the other hand, Fig.~\ref{mixd} {\bf (b)} shows how such input states are mapped by the two extreme cases of quantum channels considered above.  It is clear that the performance of MEMS as entanglement teleportation channels is qualitatively the same for mixed target states as for pure ones. When both channels are of the large-entanglement family, performances are comparatively better than under the usage of two $\rho^2$.

\section{Experimental set-up}
\label{exp}

\begin{figure}[t]
\psfig{figure=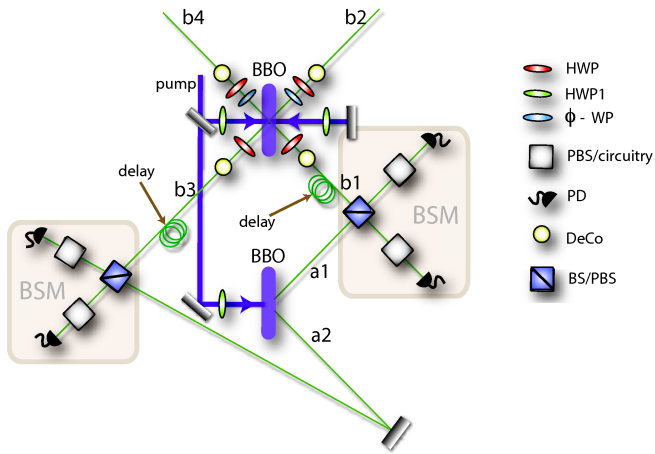,width=8.6cm,height=6.cm}
\caption{(Color Online) Proposed experimental set-up for the linear-optics implementation of the entanglement teleportation scheme via MEMS. Three pairs of entangled photons are generated via single-pass and double-pass SPDC processes. Two of such pairs embody the MEMS channels (modes $b_1\&b_2$ and $b_3\&b_4$), while the remaining pair encodes the entanglement we wish teleport. Two Bell-state measurement (BSM) stages are included in the set-up. We show the symbols for half-wave plates (HWP's) and $\phi$-wave plates (QWP's) as well as polarization-sensitive and insensitive beam splitters (PBS's and BS respectively) photo-detectors (PD's) and decoherers (DeCo)~\cite{peters} required for MEMS generation. The white cubes in each BSM stage embody a normal PBS (some optical circuitry), to be used together with BS's (PBS's) interfering modes $a_2\&b_3$ and $a_1\&b_1$ for projections onto $\ket{\Psi^-}$ ($\ket{\Phi^+}$).}
\label{setup}
\end{figure}

We now pass to discuss an experimental setting where the scheme discussed here can be probabilistically implemented with current state-of-the-art linear-optics technology. The set-up is schematically shown in Fig.~\ref{setup}. It involves six entangled photon pairs, each produced by a type-I spontaneous parametric down conversion (SPDC) process resulting from pumping a $\beta$-Barium Borate (BBO) crystal with an intense (ultra-violet) pump field. An economic and compact experimental configuration can take advantage of a double-pass scheme for the generation of the two required entangled channels~\cite{walter}, as shown in Fig.~\ref{setup}. Information would be encoded into the polarization degrees of freedom of the photons. The double-pass SPDC stage would be responsible for the generation of two entangled pairs, each involving spatial modes $b_1\&b_2$ and $b_3\&b_4$ respectively. The entanglement within each pair would be adjusted by the proper setting of input half-wave plates HWP1. Following the ``Procrustean" experimental scheme of Ref.~\cite{peters}, a sequence of linear optics elements including a HWP, a $\phi$-wave plate ($\phi$-WP) and a decoherer (DeCo) can be placed along such spatial modes in order to generate MEMS states of desired entanglement and mixedness. The single-pass SPDC stage, on the other hand, produces the qubit-pair $a_1-a_2$ whose entanglement (quantitatively set by the adjustment of a respective HWP1) we want to teleport. Our linear-optics implementation exploits the handiness of rigidly projecting pairs of photons onto $\ket{\Phi^+}$ via the use of a simple polarizing beam splitter (PBS) and some additional optical elements (indicated as ``circuitry" in Fig.~\ref{setup})~\cite{pan1}. The experimental convenience of rigid Bell measurements motivates further, a posteriori and pragmatically, the restriction we have invoked over the capabilities of Alice of projecting at will. 
The efficiency of the BSM's can be improved by using fiber-BS's, which optimize the spatial mode-matching and, overall, the performance of the set-up. Clearly, any other Bell-state discrimination stage can be adopted in order to replace the one suggested here.  For instance, by replacing the PBS-circuitry configuration mentioned above with the cascade of a beam splitter (BS), two PBS's and two photodetectors (PD's) in the Bell-state analyzer configuration of Refs.~\cite{jennewein,prevedel} (which is optimal for linear optics~\cite{nogo}), projections onto $\ket{\Psi^\pm}$ can be performed. The fact that only four-photon coincidences at the four PD's are required for the performance of the protocol makes it non-demanding in terms of data-acquisition rate. Adding a stage of state-verification after the teleportation adds, quite obviously, complexity to the set-up.

 All the ingredients of our proposal have been experimentally demonstrated, some of them being currently routinely implemented in various linear-optics labs. As a remarkable example of the realistic possibility to manipulate three SPDC processes complemented by two BSM's, we mention the multistage entanglement swapping scheme and the teleportation schemes of Ref.~\cite{pan1}. 

\section{conclusions}
\label{conclusion}
We have examined the ability to teleport entanglement through noisy quantum channels belonging to the special class of MEMS. Such an ability is not only heavily dependent on the quality of the channel being used, but also on the type of target state to be teleported. Within the validity and limits of our quantitative study, if the target state has a large overlap with $\ket{\Psi^{+}}$, the teleported state is entangled whenever the input state is, although the degree of entanglement can be easily affected. Also, differently from previous works, the entanglement was found not to be a fragile commodity. In fact, under proper circumstances, fidelity between input and teleported state appears to grow with the target entanglement. All this strongly suggests that the performance of teleportation channels can be dependent on the entanglement measure being taken,which not only quantitatively determines the amount of teleported entanglement but, in this case, shapes the density matrices of the channels as well. We expect our results to be appealing both theoretically (particularly in relation to the investigation of the role that MEMS pay in quantum protocols for information processing) and pragmatically. In fact, we have shown that an experimental demonstration of the protocol and features discussed here is well in order with state-of-the-art six-photon linear-optics technology. Requiring only experimentally non-demanding four-fold coincidences, our proposed experiment may be promptly and efficiently realized. We therefore hope that our investigation will rise the interest of the theoretical and experimental communities alike. 
\acknowledgments 
We thank R. Prevedel for invaluable discussions. We acknowledge financial support from DEL and the UK EPSRC (EP/G004579/1).


\begin{thebibliography}{99}
\bibitem{comment} We have $\hat{A}^1_{b_k}=\begin{pmatrix}\frac{r}{2}&0\\0&0\end{pmatrix},\hat{B}^1_{b_k}=\begin{pmatrix}0&\frac{r}{2}\\0&0\end{pmatrix},\hat{C}^1_{b_k}=\begin{pmatrix}0&0\\\frac{r}{2}&0\end{pmatrix}$ and $\hat{D}^1_{b_k}=\begin{pmatrix}1-r&0\\0&\frac{r}{2}\end{pmatrix}$. On the other hand, for the $\rho^2$ family of MEMS we have $\hat{A}^2_{b_k}=\begin{pmatrix}\frac{1}{3}&0\\0&0\end{pmatrix},\hat{B}^2_{b_k}=\begin{pmatrix}0&\frac{r}{2}\\0&0\end{pmatrix},\hat{C}^2_{b_k}=\begin{pmatrix}0&0\\\frac{r}{2}&0\end{pmatrix}$ and $\hat{D}^2_{b_k}=\begin{pmatrix}\frac{1}{3}&0\\0&\frac{1}{3}\end{pmatrix}$.
\end{thebibliography}
\end{document}